\documentclass[twocolumn,twoside,slac_two]{revtex4}
\usepackage{graphicx}
\usepackage{fancyhdr}
\begin{document}

\title{The LHCb detector}

\author{Eddy Jans}
\email{eddy@nikhef.nl}
\affiliation{Nikhef, P.O. Box 41882, 1009 DB Amsterdam, The Netherlands\\
\rm{on behalf of the LHCb collaboration}}

\begin{abstract}
The LHCb spectrometer is designed with special emphasis on the discovery potential of b and c-physics studies.
The main characteristics of the setup are discussed shortly.
Since the completion of the LHCb detector, it has extensively been commissioned with cosmic rays, simulated data, beam and beam-induced events.
Some commissioning results and an outlook to early physics are presented.
\end{abstract}

\maketitle

\thispagestyle{fancy}


\section{Introduction}
\begin{figure}[h]
\centering
\includegraphics[width=80mm]{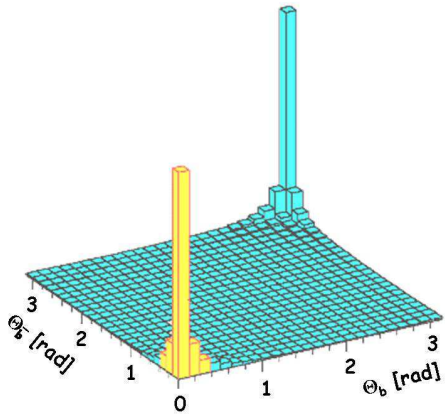}
\caption{Pythia simulation of the b$\bar \mathrm{b}$-production cross section as a function of the polar angles of the b and $\bar \mathrm{b}$ produced in the primary vertex} \label{bbarangcor}
\end{figure}

LHCb is a dedicated b and c-physics experiment at the Large Hadron Collider (LHC)
that will search for New Physics (NP) beyond the Standard Model (SM) via high-precision measurements of 
CP-violating observables and rare decays of beauty and charm-flavoured hadrons.
In the proton-proton collisions at the design energy of LHC, i.e. $\sqrt{s}$=14 TeV, the full B-hadron and B-baryon spectrum will be produced: 
$B^{0}$, $B^{\pm}$, $B_{s}^{0}$, $B_{c}^{\pm}$ and $\Lambda_{b}^{0}$. 
In Fig.~\ref{bbarangcor} the b$\bar \mathrm{b}$-production cross section is plotted as a function of the polar angles of the produced b and $\bar \mathrm{b}$ with respect to the beam direction (z-axis).
The strong angular correlation has led to the design of the LHCb detector as a forward angle spectrometer.
\section{Detector layout and specifications}
\begin{figure}[h]
\centering
\includegraphics[width=80mm]{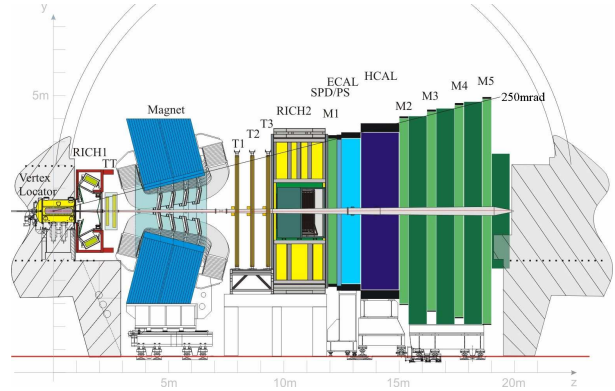}
\caption{Side view of the LHCb detector with the following components: Vertex Locator (VELO), RICH1, Trigger Turicencis (TT), dipole magnet, 
three tracking stations (T1, T2, T3), RICH2, Scintillating Pad Detector (SPD), PreShower (PS), Electromagnetic (ECAL) and Hadronic (HCAL) calorimeters and a Muon system (M1 up to M5). } \label{LHCbdetector}
\end{figure}

LHCb is a single-arm spectrometer with an angular coverage from $\sim$15 mrad to 300 (250) mrad in the bending (non-bending) plane.
The layout of the spectrometer is shown in Fig.~\ref{LHCbdetector}.
It is described in great detail in \cite{JINSTpaper}.

The key characteristics of LHCb are:
\begin{itemize}
\item Excellent vertex and proper time resolution;
\item Good particle identification (PID) properties, especially $\pi$/$K$ separation;
\item Good momentum resolution, which together with the precise direction determination by the VELO, implies precise invariant mass resolution;
\item An efficient and flexible trigger system;
\end{itemize}
These four characteristic items are discussed below.
\subsection{Vertexing}
The VELO consists of 2 halves that have to be retracted by 29~mm before the beams can be injected into LHC. Each half contains 21 modules, each of which has two silicon half disks with strips in R and $\Phi$ geometry, repectively.
The detector volume is separated from the beam volume by a 0.3~mm thick corrugated Aluminium foil.
In the closed position the innermost strip approaches the beam to 8.2~mm.
Simulations show that the expected impact parameter resolution amounts to $\delta$IP $\sim$ 14~$\mu$m + 35~$\mu$m/$p_T$,  where $p_T$ is the transverse momentum in GeV/c with respect to z.
The resolution with which the $z$ coordinate of the primary vertex can be determined is expected to be 47 $\mu$m.
For the $B_s$-decay vertex of $B_s\rightarrow$$D_{s}^{\pm}K^{\mp}$ the resolution amounts to 144 $\mu$m.
These values should be compared to the average flight path of $\sim$9~mm between the primary and secondary vertex of a proton-proton collision in which a $B_s$-meson is produced.
\subsection{Particle identification}
\begin{figure}[h]
\centering
\includegraphics[width=80mm]{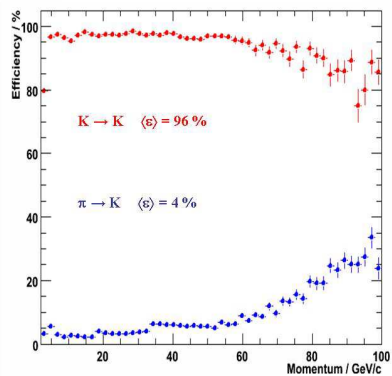}
\caption{Probability for kaons to be indentified as a kaon (red) and pions to be misidentified as a kaon (blue) as a function of momentum.} \label{pikaonidentification}
\end{figure}

Good $\pi$/$K$ seperation is a prerequisite for a large part of the experimental program of LHCb.
Therefore there are 2 RICH detectors to cover the required momentum range. RICH1 is situated in front of the magnet and uses aerogel and $C_4$$F_{10}$ radiators for the momentum range up to $\sim$60 GeV/c. The momentum range between $\sim$15 and 100 GeV/c is covered by RICH2 using a $CF_4$ radiator.
The expected kaon identification and pion misidentification probabilities are plotted in Fig.~\ref{pikaonidentification}.

The LHCb calorimeter system, that consists of PS, SPD, ECAL and HCAL, performs several tasks.
It selects hadron, electron and photon candidates with sufficient transverse energy for the first level (L0) trigger. 
Moreover, it provides particle identification as well as energy and position determination.

Muon triggering and offline muon identification are key characteristics for B-physics studies, as muons are present in many final states of $B$-decays, and as such crucial for many CP-asymmetry measurements \cite{Blusk} and rare $B$-decays \cite{Bettler}.
\subsection{Momentum resolution}
Tracking is performed by combining the information from the VELO, TT, IT and OT. The integrated field of the normal conductive dipole magnet is of the order of 4~Tm.
The expected momentum resolution varies as a function of the track momentum from 0.4 to 0.6~\%.
\subsection{Trigger}
At the design luminosity of 2$\cdot$10$^{32}$ cm$^{-2}$s$^{-1}$ the rate of events with at least two particles in the LHCb acceptance will be $\sim$10 MHz. 
The task of the trigger system is to reduce this rate by a factor 5000 to 2~kHz, which is subsequently written to storage for off-line analysis.  
Trigger decisions are based on characteristic features of the reactions under study, like the fact that b-hadrons live long, resulting in well-seperated primary and secondary vertices, and have a relatively large mass, resulting in decay products with large $p_T$.
The trigger system consists of 2 parts: Level0 (L0) which is implemented in custom-made electronics and a High Level Trigger (HLT) implemented in software and presently running on $\sim$4000 commercially available processors.
The decision of the L0 is based on the information of the PileUp detector, which identifies bunch crossings with multiple interactions, and of the calorimeter and muon systems.
The HLT first confirms and refines the high-$p_T$ L0-candidates, and then selects events with partially or fully reconstructed $B$-decay modes.
\subsection{Performance}
\begin{figure}[h]
\centering
\includegraphics[width=80mm]{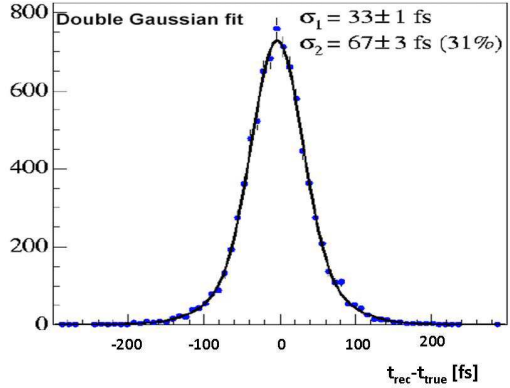}
\caption{Difference between the reconstructed and generated proper time of $B_s$ decaying into $D_{s}K$.} \label{decaytime}
\end{figure}

Many simulations have been performed to predict the response of the detector and to prepare for physics analyses \cite{MCsimulations}.
For time-dependent CP-asymmetry measurements with $B_s$-mesons it is mandatory to provide for an excellent proper time resolution in order to be able to resolve the fast $B_s$$\bar{B_s}$-oscillations.
In Fig.~\ref{decaytime} the distribution of the difference between the reconstructed and simulated $B_s$-meson proper time is plotted.
When fitted with a double-Gaussian a proper time resolution of $\sim$40 fs emerges.
This is more than sufficient to determine time dependent CP-violation in the $B_s$-system.

\begin{figure}[h]
\centering
\includegraphics[width=80mm]{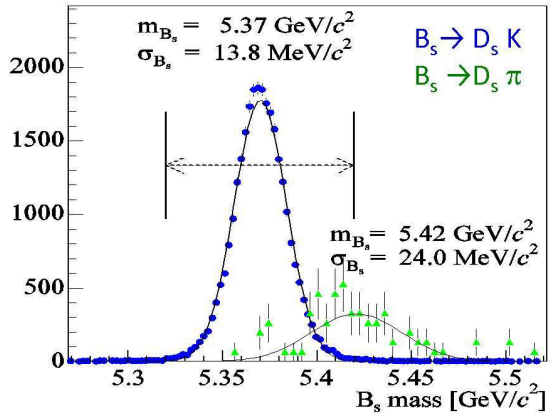}
\caption{Invariant mass distribution of the reconstructed ${B_s}$ mesons decaying into $D_{s}^{\pm}K^{\mp}$ after application of the offline cuts. Also shown is the remaining contribution from the $B_s$$\rightarrow$$D_{s}^{\pm}{\pi}^{\mp}$ decays, which is expected to be the dominant background.} \label{massresolution}
\end{figure}

Another important aspect of LHCb is its mass resolution.
In Fig.~\ref{massresolution} the invariant mass spectrum of the reaction $B_s\rightarrow$$D_{s}^{\pm}K^{\mp}$ is shown, as obtained after off-line selection, featuring an expected mass resolution of 14~MeV.
Also shown is the contribution of $B_s$$\rightarrow$$D_s\pi$, which is expected to be the dominant background, but giving only a minor contribution thanks to the excellent p.i.d. properties of LHCb.

\section{Commissioning}
Commissioning of the LHCb detector started in 2007 with individual sub-detectors being powered and read out. 
Control and tuning software were further developed until in summer 2008 all sub-detectors were operated under central control and the full chain of data taking and storage was exercised for the first time.
In August 2009 the full detector was read out at a rate of 1~MHz, very close to its design specification.
The back-end part of the DAQ system, consisting of HLT processor farm, storage, run control and monitoring, is regularly tested by means of injection of Monte Carlo events.
In the following three sub-sections some details of the commissioning are presented; those obtained with cosmic rays, beam and beam-induced events, respectively.
\subsection{Commissioning with cosmics}
\begin{figure}[h]
\centering
\includegraphics[width=80mm]{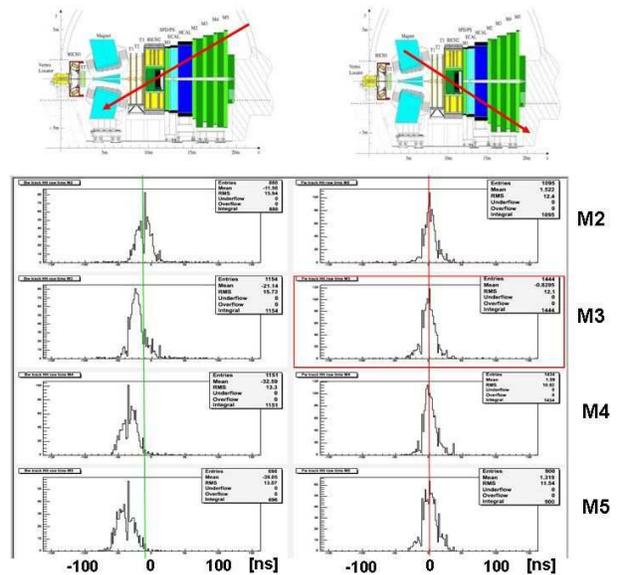}
\caption{Time difference distributions between an external trigger and muon stations M2, M3, M4 and M5, for backward (left) and forward (right) tracks.} \label{muontimealignment}
\end{figure}

Although the configuration of LHCb is not well suited for measuring cosmic rays, a few million cosmic events have been collected in the course of time.
The data are used for various calibration purposes. 
For instance the various parts of the muon system have been time aligned and its efficiency studied. 
In Fig.~\ref{muontimealignment} the time distributions with respect to the trigger, as provided by the SPD, are shown for muon stations M2-M5.
Forward tracks come from the "good" direction and are used to synchronize the muon stations.
The difference in time of flight between backward and forward tracks is clearly visible in the two sets of plots. 
Moreover, analysis code for space alignment of various sub-detectors is developed and applied to these cosmics data.
Unfortunately cosmics can not be used by the sub-detectors that cover considerably smaller areas, like IT, TT and VELO.
\subsection{Commissioning with beam}
\begin{figure}[h]
\centering
\includegraphics[width=80mm]{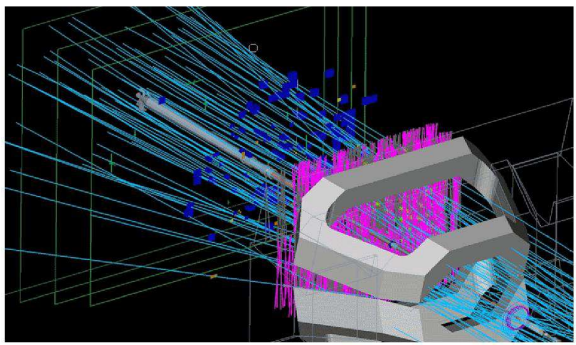}
\caption{Event display of a splash event recorded on September 10, 2008. Visible are the coils of the magnet (grey), hits in the straws of the OT (purple), hits in the CALO (dark blue) and reconstructed tracks (light blue).} \label{splashevent}
\end{figure}

On September 10, 2008 both beams of LHC circulated for the first time through the ring. 
Some detectors of LHCb were switched on when beam2, i.e. the clock-wise one, passed through Point~8.
They recorded some clean events, but also a handful of splash events, probably due to (part of) the beam hitting a collimator and generating a lot of secondary particles.
In Fig.~\ref{splashevent} an event display of such a splash event is shown.
Detector elements of the OT and CALO that registered a hit are highlighted, as well as the straight tracks that could be fitted through them.
\subsection{Commissioning with beam-induced events}
\begin{figure}[h]
\centering
\includegraphics[width=80mm]{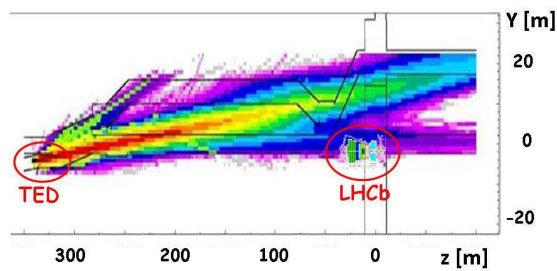}
\caption{Profile in (y,z) coordinates of the shower of particles produced by a 450~GeV proton bunch of SPS impinging on the TED, located 340~m downstream of LHCb.} \label{tedprofile}
\end{figure}

At various occasions the LHC machine group has performed injection tests in which proton bunches from SPS are directed on a beam stopper (TED) located about 340~m downstream of LHCb.
The simulated profile of emerging secondary and tertiary particles, mainly being muons, is shown in Fig.~\ref{tedprofile}.
An SPS bunch containing 5$\cdot$10$^9$ protons generates on average 7 tracks in the VELO, i.e. in an effective surface of 54~cm$^2$.
These beam-induced events are used to measure pulse shapes of the Beetle frontend chips and to synchronize the readout of the silicon detectors, both internally and with respect to other sub-detectors.
With a very limited data set consisting of only 100 bunches it turned out to be possible to time align the VELO sensors within $\sim$2 ~ns.
Also first spatial alignment results have been obtained with these so-called TED-data.
For instance the VELO module-to-module alignment parameters have been determined in situ.
Comparison with those extracted from earlier metrology measurements indicates that the module alignment parameters are consistent within 10~$\mu$m for translation and 200~$\mu$rad for rotation.
The distance between the two VELO-halves can be extracted from tracks passing through both sides.
An actual movement of 450~$\mu$m is reconstructed as 445$\pm$10~$\mu$m.
Moreover, the first alignment between VELO and TT has been obtained by means of extrapolated VELO-tracks, resulting in a residuals distribution with a $\sigma$ of 500~$\mu$m.
\section{Early physics}
The plans for early physics will evolve in the course of time as the beam energy and luminosity delivered by LHC will become more clear.
Notwithstanding these uncertainties a series of topics has been identified that can already lead to tantalizing physics results with only limited integrated luminosity.
Of these the following have been presented at this conference:
\begin{itemize}
\item inclusive V0 and $D$-meson production,\\
this analysis will be based on a 100~M minimum bias sample and 1~M $\mu$($\mu$)-events.
Using reconstructed J/$\psi$$\rightarrow$$\mu$$^+$$\mu$$^-$ decays both the prompt J/$\psi$ and b$\rightarrow$J/$\psi$ production cross section will be determined in the pseudo-rapidity range 2$\leq$$\eta$$\leq$5. 
Strangeness production and hadronization will be investigated via the ratio ${\overline{\Lambda}}$${/}$${\Lambda}$ as a function of $p_T$ and $\eta$.
See \cite{Dettori}.
\item search for New Physics in rare decays,\\
rare decays like $B_{s}^{0}$$\rightarrow$$\mu$$^+$$\mu$$^-$ are flavour changing neutral current decays that are forbidden at the tree level.
Determination of the branching fraction of this rare decay opens a venue for discovering New Physics, due to the possible contribution of virtual particles in the box and penguin diagrams. 
Background rejection is a major issue in these studies, but is expected to be manageable thanks to the good vertexing capabilities and mass resolution of LHCb.
See \cite{Bettler}.
\item searching for New Physics in CP-violation,\\
predictions of the Standard Model will, among others, be tested via the determination of the $B_s$-mixing phase $\phi$$_s$ from a flavour tagged, angular analysis of $B_{s}^{0}$$\rightarrow$J/$\psi$ $\phi$. 
See \cite{Blusk}.
\end{itemize}
\section{Conclusions and outlook}
Since the construction of the LHCb detector is completed, the setup has extensively been commissioned with cosmic rays, simulated data, beam and beam-induced events.
These data permitted sub-detectors to perform internal, as well as mutual, time alignment.
First spatial alignment constants have been determined showing only small deviations from earlier metrology results.
The detector frontend electronics has been read out at 1~MHz.
The LHCb collaboration is looking forward to the first proton-proton collisions at 3.5~TeV per beam and their physics outcome in this terra incognita of particle physics.
\bigskip 

\end{document}